\documentclass{IEEEtran}
\usepackage{amsmath, amssymb, graphicx}
\usepackage{algorithmic}
\usepackage{algorithm}
\usepackage{lettrine} % for drop caps
\usepackage{array} % required for text wrapping in tables
\usepackage{refcount}
\usepackage{flushend}

\title{Effect of Phase Shift Errors on the Security of UAV-assisted STAR-RIS  IoT Networks}

\author{Mustafa Gusaibat, Mohammed Hnaish, Abdelhamid Salem, \textit{Member, IEEE}, Khaled Rabie, \textit{Senior Member, IEEE}, Zubair Md Fadlullah, \textit{Senior Member, IEEE}, Wali Ullah Khan, Mohamad A. Alawad, and Yazeed Alkhrijah

\thanks{Corresponding authors: A. Salem and Mohamad A. Alawad.}

\thanks{Mustafa Gusaibat and Mohammed Hnaish are with the Department of Electronic
 and Electrical Engineering, Benghazi University, Benghazi 1308, Libya, (emails: mustafa.a.gusiebat@gmail.com, mohammedhnaish@gmail.com).}

\thanks{Abdelhamid Salem is with the Department of Electronic and Electrical Engineering, University College London, WC1E 6BT London, U.K., and also with the Department of Electronic and Electrical Engineering, Benghazi University, Benghazi, Libya (e-mail: a.salem@ucl.ac.uk).}

\thanks{Khaled Rabie is with the Department of Computer Engineering, King Fahd University of Petroleum and Minerals (KFUPM), Dhahran, Saudi Arabia (email: k.rabie@kfupm.edu.sa).}

\thanks{Zubair Md Fadlullah is with Department of Computer Science, Western University, London Ontario N6G 2V4, Canada (e-mail: zfadlullah@ieee.org).}

\thanks{Wali Ullah Khan is with the Interdisciplinary Centre for Security,
Reliability, and Trust, University of Luxembourg, 1855 Luxembourg City,
Luxembourg (e-mail: waliullah.khan@uni.lu).}

\thanks{Mohamad A. Alawad and Yazeed Alkhrijah are with the Department of Electrical Engineering, Imam Mohammad Ibn Saud Islamic University (IMSIU), Riyadh, Saudi Arabia, (emails: maawaad@imamu.edu.sa, ymalkhrijah@imamu.edu.sa).}

}

\date{}

\begin{document}

\maketitle
\begin{abstract}
Unmanned aerial vehicles (UAV)-mounted simultaneous transmitting and reflecting reconfigurable intelligent surface (STAR-RIS) systems can provide full-dimensional coverage and flexible deployment opportunities in future 6G-enabled IoT networks. However, practical imperfections such as jittering and airflow of UAV could affect the phase shift of STAR-RIS, and consequently degrade network security. In this respect, this paper investigates the impact of phase shift errors on the secrecy performance of 
UAV-mounted  STAR-RIS-assisted IoT systems.  More specifically, we consider a UAV-mounted STAR-RIS-assisted non-orthogonal multiple access (NOMA) system where IoT devices are grouped into two groups: one group on each side of the
STAR-RIS. The nodes in each group are considered as potential
Malicious nodes for the ones on the other side. By modeling phase
estimation errors using a von Mises distribution, an analytical closed-form
expressions for the ergodic secrecy rates 
under imperfect phase adjustment are derived. An optimization problem to maximize the weighted
sum secrecy rate (WSSR) by optimizing the UAV placement is formulated and is then solved using a linear grid-based algorithm. Monte Carlo simulations are provided to validate
the analytical derivations. The impact of phase
estimation errors on system’s secrecy performance is analyzed, providing critical insights for the practical realisation of STAR-RIS deployments for secure
UAV-enabled IoT networks. 
\end{abstract}

\begin{IEEEkeywords}
Unmanned aerial vehicles, STAR-RIS, physical layer security, phase shift errors, NOMA, secrecy performance. 
\end{IEEEkeywords}

\section{Introduction}

\lettrine{U}{nmanned} aerial vehicles (UAVs) have become increasingly vital for future wireless networks due to their autonomous operation, operational flexibility, and cost-efficient deployment \cite{ref1}. These systems enable robust and reliable communications, particularly for rapid infrastructure establishment and emergency scenarios. However, UAV-assisted communications face two fundamental challenges: i) inherent security vulnerabilities arising from their broadcast nature and the domination of line-of-sight (LoS) channels, and ii) the operational limitations caused by constrained energy resources and impractical battery replacement requirements. Addressing these interdependent challenges of security enhancement and energy optimization represents a critical research direction for UAV-enabled wireless systems.

Furthermore, reconfigurable intelligent surfaces (RIS) offers a transformative approach for IoT networks, capable of intelligently manipulating signals through dynamic control of reflection amplitudes and phase shifts \cite{ref2}. Due to their compact size, light weight design, and cost-effectiveness, RIS can be seamlessly integrated onto UAVs. Leveraging RIS-enabled programmable propagation environments offers significant potential to enhance coverage, physical-layer security and energy efficiency of UAV-assisted IoT networks \cite{ref3}, \cite{ref4}. Specifically, RIS provides an efficient solution compared to traditional multi-antenna systems without requiring additional energy-intensive active components \cite{ref3}. Furthermore, RIS enables dynamic reconfiguration of wireless channels to improve link quality of intended nodes receivers while degrading undesired nodes links \cite{ref4}.

%Furthermore, reconfigurable intelligent surfaces (RIS) represent a transformative technology for the sixth-generation (6G) communication, capable of intelligently manipulating radio environments through dynamic control of reflection amplitudes and phase shifts \cite{ref2}. Due to their compact size, lightweight design, and cost-effectiveness, RIS can be seamlessly integrated onto UAVs. Leveraging RIS-enabled programmable propagation environments offers significant potential to enhance  secrecy and energy efficiency of UAV-assisted communications \cite{ref3}, \cite{ref4}. Specifically, RIS provides an efficient solution compared to traditional multi-antenna systems or active relay-based UAVs \cite{ref3}. Furthermore, RIS enables dynamic reconfiguration of wireless channels to improve the quality of signals at intended receivers while degrading undesired receivers channels \cite{ref4}. 

In contrast to conventional passive RISs that only reflect signals, simultaneously transmitting and reflecting RIS (STAR-RIS) dynamically manipulates both transmitted and reflected signals via independent phase-shift control, enabling full-dimensional wireless reconfiguration and simultaneous service for both sides of the surface \cite{ref5}, \cite{ref6}. STAR-RIS operation typically follows three fundamental protocols: mode switching (MS), where RIS elements are divided into two distinct groups, one for reflection and the other for transmission; energy splitting (ES) where all elements processes both reflected and transmitted signals; and time switching (TS) elements experience a dynamic switching between reflection and transmission modes over different time slots.

Mounting STAR-RIS on UAVs further extends these benefits by offering flexible deployment, improved line-of-sight (LoS) connectivity, and dynamic beamforming capabilities \cite{16,20}. Such integrated platforms are well-suited for environments with obstructed links, where they can improve both coverage and physical layer security (PLS) \cite{18,19}. By exploiting the UAV’s mobility and the STAR-RIS’s full-space wave control, these systems enable adaptive and secure links, especially in non-LoS scenarios involving eavesdroppers. However, practical issues like phase shift errors caused by UAV vibrations and aerodynamic disturbances can significantly impair performance, underscoring the need for robust design and optimization strategies \cite{36,39}.

%\subsection{Prior Works}

Several prior works have explored the integration of STAR-RIS and UAV systems in the context of PLS and NOMA-enabled communication. Zhang \textit{et al.} \cite{zhang2022star} analyzed the secrecy outage performance of STAR-RIS-assisted uplink NOMA systems and proposed an alternating optimization algorithm to enhance secrecy rates by jointly tuning transmit power, beamforming, and STAR-RIS coefficients. Li \textit{et al.} \cite{li2023secrecy} extended this by deriving closed-form secrecy expressions and showed how STAR-RIS improves secrecy compared to traditional RIS.% In \cite{han2022an}, Han \textit{et al.} introduced artificial noise (AN) to further enhance security in STAR-RIS-based NOMA, optimizing under quality of service (QoS) constraints.
Tang \textit{et al.} \cite{tang2021ris} addressed the problem of passive RISs degrading secrecy at high element counts by proposing a new RIS design that nullifies the eavesdropper's channel. Wang \textit{et al.} \cite{wang2025covert} examined covert communications in UAV-ground STAR-RIS NOMA networks, optimizing UAV placement and STAR-RIS settings to maximize covert throughput while ensuring detectability constraints. Salem \textit{et al.} \cite{salem2024phase} specifically analyzed the impact of phase-shift errors in RIS-assisted systems, revealing that active RIS configurations offer resilience and improved secrecy in power-constrained uplink scenarios. Similarly, Li \textit{et al.} \cite{li2024active} investigated active RIS-assisted NOMA networks, providing closed-form secrecy outage expressions and demonstrating the superiority of active RIS (ARIS) in both perfect and imperfect successive interference cancellation (SIC) regimes. %Together, these works form a strong foundation for studying the practical challenges and optimization strategies in UAV-mounted STAR-RIS systems under realistic imperfections such as phase shift errors.

%\subsection{Motivation and Contributions}

 Practical implementations of UAV-mounted STAR-RIS systems encounter several technical challenges. First, phase errors induced by UAV mobility—such as jittering and airflow disturbances—can significantly distort phase alignment, degrading overall secrecy performance~\cite{8}. Second, optimizing STAR-RIS configurations under secrecy and energy efficiency constraints remains complex, especially under payload limitations~\cite{9}. Third, joint optimization of UAV placement and RIS phase shifts becomes computationally intensive in the presence of multiple eavesdroppers with unknown locations~\cite{10}.

Motivated by these challenges, this paper investigates the impact of phase shift errors on the secrecy performance of UAV-mounted STAR-RIS-assisted NOMA IoT network in uplink scenarios. A comprehensive system model is developed, incorporating several realistic impairments and novel analytical tools.

The main contributions of this work are summarized as follows:

\begin{itemize}

    \item  We consider a STAR-RIS-assisted uplink NOMA IoT network, where a UAV-mounted STAR-RIS serves as a passive relay between access point (AP) and two clusters of  IoT devices located on its transmission and reflection sides. In each time slot, one node from each cluster transmits using NOMA pairing while the remaining idle nodes act as potential malicious nodes. The signal-to-noise ratio (SNR) distribution is analyzed under combined effects of Nakagami fading and phase estimation error. Closed-form expressions for the ergodic secrecy rate are derived using Laguerre quadrature, offering high computational efficiency and analytical precision compared to simulation-based methods.

    \item  The weighted sum secrecy rate (WSSR) maximization problem is formulated. A linear grid search algorithm optimizes UAV placement, balancing accuracy and computational complexity is proposed.
    \begin{figure}[tbp]
    \vspace{-10pt} % Reduce space above
    \centering
    \includegraphics[width=0.4\textwidth]{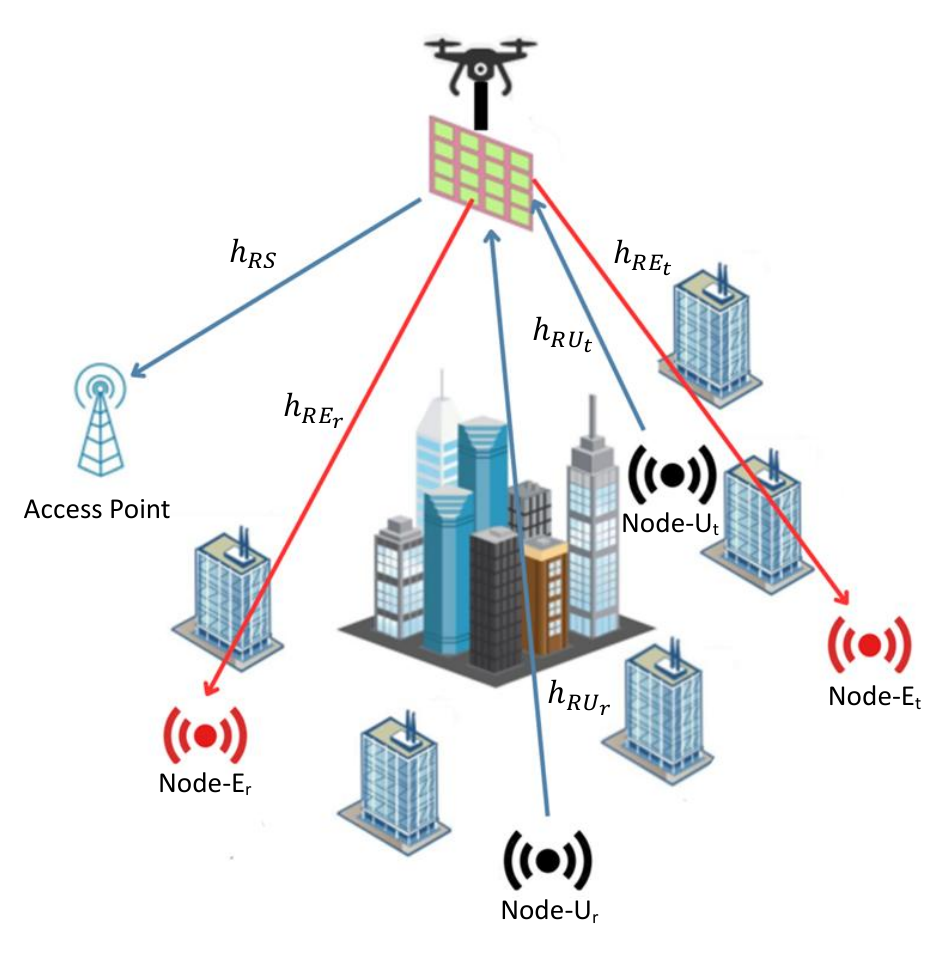}
    \vspace{-10pt} % Adjust this value as needed
    \caption{UAV-mounted STAR-RIS-assisted secure uplink NOMA for IoT network.}
    \label{fig:system}
    %\vspace{20pt} % Reduce space below
    \end{figure}
    \item  Extensive simulations validate the analytical findings and proposed optimization framework, revealing the sensitivity of secrecy performance to phase error and demonstrating the benefits of the UAV placement design.
    \item The results of this paper show the impact of concentration parameter values, the number of reflected elements and the UAV placement on secrecy performance where higher values of concentration parameters lead to less phase error and better secrecy performance, and the amount of signal power received by each node depends on the number of reflected elements. In addition, UAV placement optimization enhances WSSR. 

\end{itemize}

This work offers both theoretical insight and practical design guidelines for secure, efficient deployment of STAR-RIS-enabled UAV communication systems in future wireless networks.

%\subsection{Organization}
%The rest of the paper is organized as follows. Section II presents the system model for the UAV-mounted STAR-RIS-assisted uplink NOMA communication, including channel modeling and phase shift error analysis. Section III derives closed-form expressions for the ergodic secrecy rate under imperfect phase estimation. Section IV proposes UAV placement optimization to maximize WSSR. Section V validates the analytical results through numerical simulations, demonstrating the impact of phase errors and optimized parameters. Finally, Section VI concludes the paper and discusses future directions.

\section{System Model}

As shown in Fig. 1, we consider a UAV-mounted STAR-RIS assisted NOMA IoT system, where an access point (AP) receives signals from nodes (IoT devices) through a UAV-mounted STAR-RIS.  nodes are divided into two clusters based on their location relative to the STAR-RIS, one cluster lies in the reflection side, and the other in the transmission side. 
In each time slot, one node from each cluster performs uplink transmission,  The AP and the nodes are equipped with a single antenna. By applying the NOMA pairing scheme, both nodes can simultaneously access the AP by sharing the same time slot and frequency. The number of reflecting elements is $M$ and the number of nodes is $K$, $K=K_r +K_t$, where $ K_r$ and $K_t$ number of nodes in the reflection and transmission regions, respectively. In this paper the number of nodes is assumed to be two nodes on each side to reduce complexity. 

Due to the broadcast nature of wireless transmissions,  the nodes that are not transmitting in the current time slot (the idle nodes both clusters) ($E_q$) may aim at wiretapping active nodes ($U_q$) data. Therefore, each non-transmitting node is considered as potential malicious node to the node from the opposite side, where $q\in \mathcal{}$$[r,t]$.

The channel coefficients between the $U_q$ node and the $m$-th reflected element, the $m$-th reflected element and the AP, and the $m$-th reflected element and the $E_{q'}$ node on the opposite side are denoted by $h_{U_{q}R_{m}}$, $h_{R_{m}S}$, and $h_{R_{m}E_{q'}}$, respectively, which are modeled as Nakagami fading channels.

In this paper, a 3D Cartesian coordinate system is adopted. The horizontal locations of nodes $k \in \mathcal{K}$  are denoted by $C_k = [x_k, y_k, z_k]$. Moreover, the locations of AP and the UAV are denoted by $C_s = [0, 0, z_s]$ and $C_u = [x_u, y_u, z_u]$, respectively.

\subsection{Signal Model}

The NOMA pairing scheme is employed to reduce system complexity by pairing two nodes from different sides of the STAR-RIS in the same time slot. Therefore, the $U_{q}$ signal is expressed as:

\begin{equation}
x_{q} = \sqrt{P_{q}} s_{q}.
\end{equation}
Signal received at the AP can be expressed as:
\begin{equation}
\begin{split}
y_{s} &= \sqrt{P_{q}L_{U_{q}{R}}L_{RS}}\left[\sum_{m=1}^{M} h_{U_{q}R_{m}} h_{R_{m}S} \sqrt{\lambda_{q}^{m}} e^{j\theta_{q}^{m}}\right] x_{q} \\
&\quad + \sqrt{P_{q'}L_{U_{q'}{R}}L_{RS}}\left[\sum_{m=1}^{M} h_{U_{q'}R_{m}} h_{R_{m}S} \sqrt{\lambda_{q'}^{m}} e^{j\theta_{q'}^{m}}\right] x_{q'} \\ &\quad+ n_{t},
\end{split}
\end{equation}
where $\lambda_{q}^{m} \in [0,1]$ and $\theta_{q}^{m} \in [0,2\pi)$ are the amplitude and phase shift of element $m$, respectively. where $q \in \{r,t\}$, $U_{q'}$ represents the active node that shares the same time slot with the node $U_{q}$ and is assumed to be from the opposite side. The STAR-RIS is assumed to be operating on ES mode, therefore, to prevent violating the law of energy conservation $\lambda_{r}^{m} + \lambda_{t}^{m} = 1$ . To reduce complexity $\theta_{r}^{m}$ and $\theta_{t}^{m}$ are assumed independent. $n_{t} \sim \mathcal{CN}(0, N_{0})$ is AWGN. However, due to imperfections caused by jittering and airflow effects, it is difficult to precisely adjust the reflecting elements, which leads to practical phase shifts with error $\theta^{{err}}$, thus:
\begin{equation}
\theta_{q}^{m} = \theta^{{opt}} + \theta^{{err}},
\end{equation}

where 
\begin{equation}
L_{U_{q}{R}} = \beta d_{U_{q}{R}}^{-a} \quad \text{and} \quad L_{RS} = \beta d_{RS}^{-a}
\end{equation}
are the large scale path-loss whereas 
\begin{equation}
d_{U_{q}{R}} = \sqrt{(x_u - x_k)^2+(y_u - y_k)^2 + (z_u - z_k)^2},
\end{equation}
and
\begin{equation}
d_{RS} = \sqrt{(x_u - x_n)^2 + (y_u - y_n)^2 + (z_u - z_n)^2},
\end{equation}
represents the distances from node to UAV and from UAV to AP respectively. furthermore, $\beta$ is propagation path loss at 1 meter distance and $a$ is the path loss exponent.

The SNR of the node ($U_{q}$) at the AP ($S$) can be given by:
\begin{equation}
\begin{split}
\gamma_{s,q} &= \frac{P_{q}L_{U_{q}{R}}L_{RS} \left|\sum_{m=1}^{M} h_{U_{q}R_{m}} h_{R_{m}S} \sqrt{\lambda_{q}^{m}} e^{j\theta_{q}^{m}}\right|^{2}}{P_{q'}L_{U_{q'}{R}}L_{RS} \left|\sum_{m=1}^{M} h_{U_{q'}R_{m}} h_{R_{m}S} \sqrt{\lambda_{q'}^{m}} e^{j\theta_{q'}^{m}}\right|^{2} + N_{0}}.
\end{split}
\end{equation}
Since NOMA is implemented, the AP performs SIC to separate signals, however, the first signal to be decoded suffers inter-device interference  $P_{q'} L_{U_{q'}{R}}L_{RS}\left|\sum_{m=1}^{M} h_{U_{q'}R_{m}} h_{R_{m}S} \sqrt{\lambda_{q'}^{m}} e^{j\theta_{q'}^{m}}\right|^{2}$ which weaks its SNR.

Thus, the SNR of the node ($U_{q'}$) is:
\begin{equation}
\gamma_{s,q'} = \frac{P_{q'}L_{U_{q'}{R}}L_{RS} \left|\sum_{m=1}^{M} h_{U_{q'}R_{m}} h_{R_{m}S} \sqrt{\lambda_{q'}^{m}} e^{j\theta_{q'}^{m}}\right|^{2}}{N_{0}}.
\end{equation}

The received signal at $E_{q}$ can be expressed as:
\begin{equation}
\begin{split}
y_{e,q} &= \sqrt{P_{q}L_{U_{q}{R}}L_{RE_{q'}}}\left[\sum_{m=1}^{M} h_{U_{q}R_{m}} h_{R_{m}E_{q'}} \sqrt{\lambda_{q'}^{m}} e^{j\theta_{q'}^{m}}\right] x_{q} + n_{t}.
\end{split}
\end{equation}

The SNRs of the node ($U_{q}$) at $E_{q'}$ and  the node ($U_{q'}$) at $E_{q}$  are given respectively as:
\begin{equation}
\begin{split}
\gamma_{e,q} &= \frac{P_{q}L_{U_{q}{R}}L_{{R}{E_{q'}}} \left|\sum_{m=1}^{M} h_{U_{q}R_{m}} h_{R_{m}E_{q}'} \sqrt{\lambda_{q'}^{m}} e^{j\theta_{q'}^{m}}\right|^{2}}{ N_{0}},
\end{split}
\end{equation}

\begin{equation}
\begin{split}
\gamma_{e,q'} &= \frac{P_{q'}L_{U_{q'}{R}}L_{{R}{E_{q}}} \left|\sum_{m=1}^{M} h_{U_{q'}R_{m}} h_{R_{m}E_{q}} \sqrt{\lambda_{q}^{m}} e^{j\theta_{q}^{m}}\right|^{2}}{ N_{0}}.
\end{split}
\end{equation}

The data rates of node ($U_{q}$) at the AP and node ($U_{q}$) at ($E_{q'}$) can be given by
\begin{equation}
R_{s,q} = \log_{2}(1 + \gamma_{s,q}),
\end{equation}

\begin{equation}
R_{e,q} = \log_{2}(1 + \gamma_{e,q}).
\end{equation}

The secrecy rate of a specific side of STAR-RIS is the difference between the data rate of the node at the AP and the maximum intercepted data rate, which is expressed as:
\begin{equation}
R_{q}^{{sec}} = \left[R_{s,q} - \max R_{e,q'}\right]^{+}.
\end{equation}

\subsection{SNR Distribution}

The SNR at AP and node $E_{q}$ is influenced by the phase shifts of the reflecting elements of the RIS. The optimal phase shift $\theta^{{opt}} $ for the $m$-th reflecting element is given by:
\begin{equation}
\theta^{{opt}} = -(\theta_{R_{m}S} + \theta_{U_{q}R_{m}}),
\end{equation}
where $\theta_{R_{m}S}$ and $\theta_{U_{q}R_{m}}$ are the phases of the channels $h_{R_{m}S}$ and $h_{U_{q}R_{m}}$, respectively.

\subsubsection{Phase Estimation Errors}

practical imperfections cause the actual phase shift $\theta_{q}^m$ deviates from the optimal phase:
\begin{equation}
\theta_{q}^m = \theta^{{opt}} + \theta^{{err}}.
\end{equation}
It should be highlighted that the phase error $\theta^{\text{err}}$ is assumed to has a von Mises distribution with zero mean and concentration parameter $\kappa$. The probability density function (PDF) and trigonometric moments of $\theta^{\text{err}}$ are given by:
\begin{equation}
f_{\theta^{\text{err}}}(x) = \frac{e^{\kappa \cos(x)}}{2\pi I_{0}(\kappa)},
\end{equation}
\begin{equation}
\varphi_{p} = \mathbb{E}\left(e^{ip\theta^{\text{err}}}\right) = \frac{I_{p}(\kappa)}{I_{0}(\kappa)},
\end{equation}
where $I_{0}(\cdot)$ and $I_{p}(\cdot)$ are the modified Bessel functions of the first kind of order zero and $p$, respectively.

\subsubsection{SNR Expressions}

With phase errors, the SNRs of node ($U_{q}$) and node ($U_{q'}$) can be expressed as:
\begin{equation}
\begin{split}
\gamma_{s,q} &= \frac{\rho_{s,q}\left|\sum_{m=1}^{M} \left|h_{U_{q}R_{m}}\right|\left| h_{R_{m}S}\right| \sqrt{\lambda_{q}^{m}} e^{j\theta^{{err}}}\right|^{2}}{\rho_{s,q'} \left|\sum_{m=1}^{M}\left| h_{U_{q'}R_{m}}\right|\left| h_{R_{m}S} \right|\sqrt{\lambda_{q'}^{m}} e^{j\theta^{{err}}}\right|^{2} + 1}, \label{eq:SNRs,q}
\end{split}
\end{equation}
\begin{equation}
\gamma_{s,q'} = \rho_{s,q'} \left|\sum_{m=1}^{M} \left|h_{U_{q'}R_{m}}\right|\left| h_{R_{m}S}\right| \sqrt{\lambda_{q'}^{m}} e^{j\theta^{\text{err}}}\right|^{2},
\end{equation}
 where \(\rho_{s,q} = \frac{P_qL_{U_{q}{R}}L_{{R}{S} }}{N_{0}}\) and  \(\rho_{s,q'} = \frac{P_{q'}L_{U_{q'}{R}}L_{{R}{S} }}{N_{0}}\).
The corresponding SNRs are written as:
\begin{equation}
\gamma_{s,q} = \frac{\rho_{s,q}A}{\rho_{s,q'}B+1},
\end{equation}
\begin{equation}
\gamma_{s,q'} = {\rho_{s,q'}B},
\end{equation}
where       
\begin{equation}
A = \left| \sum_{m=1}^{M} |h_{U_{q}R_{m}}| |h_{R_{m}S}| \sqrt{\lambda_{q}^{m}} e^{j \theta^{err}_m} \right|^{2}, \label{eq:eqA}
\end{equation}        
\begin{equation}
B = \left| \sum_{m=1}^{M} |h_{U_{q'} R_m}| |h_{R_m S}| \sqrt{\lambda_{q'}^m} e^{j\theta^{err}_m} \right|^2 ,
\end{equation}
     with \(P_{q}\) and \(P_{q'}\) represents transmit power of \(U_{q}\) and \(U_{q'}\), respectively,  and \(N_{0}\) representing the variance of AWGN.
By denoting:
\begin{equation}
A = \left| \sum_{m=1}^{M} |h_{U_{q}R_{m}}| |h_{R_{m}S}| \sqrt{\lambda_{q}^{m}} e^{j \theta^{err}_m} \right|^{2} = \left|U + jV\right|^{2},
\end{equation}
the expectations and variances of  $U$ (derivations are given in Appendix A, Appendix B) can be obtained as:
\begin{equation}
\begin{split}
\mathbb{E}(U) &= M\sqrt{\lambda_{q}^{m}} \mathbb{E}(|h_{U_{q}R_{m}}| |h_{R_{m}S}| \cos \theta^{\text{err}}_m) \\
&= M\sqrt{\lambda_{q}^{m}} \alpha^{2} \varphi_{1},
\end{split}
\end{equation}
\begin{equation}
\begin{aligned}
\mathbb{V}(U) &= M \mathbb{E}(\sqrt{\lambda_{q}^{m}}|h_{U_{q}R_{m}}| |h_{R_{m}S}| \cos \theta^{{err}}_m)^{2} \\
&\quad - M \mathbb{E}^{2}(\sqrt{\lambda_{q}^{m}}|h_{U_{q}R_{m}}| |h_{R_{m}S}| \cos \theta^{\text{err}}_m) \\
&= \frac{M \lambda_{q}^{m}}{2} (1 + \varphi_{2} - 2 \alpha^{4} \varphi_{1}^{2}),
\end{aligned}
\end{equation}

where
\begin{equation}
\alpha = \sqrt{\mathbb{E}(|h_{U_{q}R_{m}}|)\mathbb{E}(|h_{R_{m}S}|)},
\end{equation}
\begin{equation}
    \quad \mathbb{E}(|h_{U_{q}R_{m}}|) = \frac{\Gamma(m_{UR} + 0.5)}{\Gamma(m_{UR})  \sqrt{m_{UR}}},
\end{equation}
\begin{equation}
\mathbb{E}(|h_{R_{m}S}|) = \frac{\Gamma(m_{RS} + 0.5)}{\Gamma(m_{RS}) \sqrt{m_{RS}}}.
\end{equation}

According to \cite{8}, the PDF of $A$ is:
\begin{equation}
f_{A}(a) = \frac{a^{m_{URS} - 1} {m_{URS}}^{m_{URS}}}{\Gamma(m_{URS}) {\Omega_{URS}}^{m_{URS}}} e^{-\frac{m_{URS}}{\Omega_{URS}} a},
\end{equation}
where:
\begin{equation}
\begin{split}
m_{URS} &= \frac{\mathbb{E}^{2}(U)}{4 \mathbb{V}(U)} = \frac{M \alpha^{4} \varphi_{1}^{2}}{2(1 + \varphi_{2} - 2 \alpha^{4} \varphi_{1}^{2})}, \\
\Omega_{URS} &= \mathbb{E}^{2}(U) = M^{2}\lambda_{q}^{m} \alpha^{4} \varphi_{1}^{2}.
\end{split}
\end{equation}

Similarly, we can obtain the distribution of $B$.

The SNRs of nodes ($U_{q}$) and ($U_{q'}$) at the eavesdroppers can be expressed as:

\begin{equation}
\begin{split}
\gamma_{e,q} &= \frac{P_{q}L_{U_{q}{R}}L_{{R}{E_{q'}}} \left|\sum_{m=1}^{M} h_{U_{q}R_{m}} h_{R_{m}E_{q'}} \sqrt{\lambda_{q'}^{m}} e^{j\theta_{q'}^{m}}\right|^{2}} { N_{0}}.
\end{split}
\end{equation}

In ideal case, $\theta^{m}$ values are designed to cancel the effects of the legitimate nodes links, which maximizes SNRs at AP, the phase shift at malicious nodes is given by: 
\begin{equation}
\phi^{m} = -\theta_{R_{m}S} + \theta_{R_{m}E_{q}} + \theta^{err},
\end{equation}
where $\phi^{m}$ is uniformly distributed over $[-\pi,\pi)$, which corresponds to complete lack of knowledge about the phase \cite{8}.
The SNR ratio is $\gamma_{e,q} = {\rho_{e,q}C},$
%\begin{equation}
%\gamma_{e,q} = {\rho_{e,q}C},
%\end{equation}
where       
\begin{equation}
C = \left| \sum_{m=1}^{M} |h_{U_{q}R_{m}}| |h_{R_{m}E_{q'}}| \sqrt{\lambda_{q'}^{m}} e^{j \phi^{m}} \right|^{2},
\end{equation}     
where \(\rho_{e,q} = \frac{P_qL_{U_{q}{R}}L_{{R}{E_{q'}} }}{N_{0}}\), Furthermore, $C$ follows an exponential distribution with mean $M$.

\section{PERFORMANCE ANALYSIS}

In this section, we derive an accurate expression for the ergodic capacity.

\subsection*{A. Ergodic Capacity}

Using \eqref{eq:SNRs,q}, the ergodic capacity of the reflection side node is expressed as
\begin{align}
C_q &= \mathbb{E}[\log_2(1 + {\gamma_{s,q}})] \notag = \mathbb{E} \left[ \log_2 \left( 1 + \frac{\rho_{s,q}A}{\rho_{s,q'}B+1} \right) \right].
\end{align}

In \cite{ref9}, a moment generating function (MGF)-based lemma for the capacity evaluation was derived as
\begin{equation}
\mathbb{E} \left[ \ln \left( 1 + \frac{X}{Y+1} \right) \right] = \int_0^\infty \frac{\mathcal{M}_{Y}(z) - \mathcal{M}_{XY}(z)}{z} e^{-z} \text{d}z,
\end{equation}
where $\mathcal{M}_{Y}(z)$ is the MGF of $Y$ and $\mathcal{M}_{XY}(z)$ is the joint MGF of $X$ and $Y$. since X and Y are independent ${M}_{XY}(z) = \mathcal{M}_{X}(z)\mathcal{M}_{Y}(z)$,
%\begin{equation}
%\mathcal{M}_{XY}(z) = \mathcal{M}_{X}(z)\mathcal{M}_{Y}(z),
%\end{equation}
and hence the expression becomes: 

\begin{equation}
\begin{aligned}
C_q &= \frac{1}{\ln(2)} \mathbb{E} \left[ \ln \left( 1 + \frac{\rho_{s,q} A}{\rho_{s,q'} B + 1} \right) \right] \\
&= \frac{1}{\ln(2)} \int_0^\infty \frac{\mathcal{M}_{Y}(z)(1 - \mathcal{M}_{X}(z))}{z} e^{-z} \text{d}z,
\label{eq:2MGF}
\end{aligned}
\end{equation}

where $X = \rho_{s,q} A,$ and $Y =\rho_{s,q'}B,$
%\begin{equation}
%X = \rho_{s,q} A,
%\end{equation}
%\begin{equation}
%Y =\rho_{s,q'}B,
%\end{equation}
\begin{equation}
\mathcal{M}_{X}(z) = \left(1+\frac{z\Omega_{URS}\rho_{s,q}}{m_{URS}}\right)^{-m_{URS}},
\end{equation}
\begin{equation}
\mathcal{M}_{Y}(z) = \left(1+\frac{z\Omega_{U'RS}\rho_{s,q '}}{m_{U'RS}}\right)^{-m_{U'RS}}.
\end{equation}
Similarly,  the ergodic capacity of the transmission side node is expressed as  
\begin{align}
C_{q'} &= \frac{1}{\ln(2)} \mathbb{E} \left[ \ln \left( 1 + {\rho_{s,q '}Y} \right) \right] \notag \\
       &= \frac{1}{\ln(2)}\int_0^\infty \frac{(1-\mathcal{M}_{Y}(z))}{z} e^{-z} \text{d}z. \label{eq:Cq'}
\end{align}
For malicious nodes capacities MGF-based lemma is used 
\begin{equation}
C_{e,q} = \frac{1}{\ln(2)} \int_0^\infty \frac{(1 - \mathcal{M}_{Y}(z))}{z} e^{-z} \text{d}z, \label{eq:Ce,q}
\end{equation}

where  $Y = \rho_{e,q} C$ , and ${M}_{Y}(z) = \frac{1}{1+M\rho_{e,q} z}$ .
%\begin{equation}
%Y = \rho_{e,q} C,
%\end{equation}
%\begin{equation}
%\mathcal{M}_{Y}(z) = \frac{1}{1+N\rho_{e,q} z}.
%\end{equation}
Similarly, we can obtain $C_{e,q'}$. 

It is notable that numerical integration can be employed to efficiently evaluate \eqref{eq:2MGF}, offering substantially faster computation compared to Monte Carlo simulations. Furthermore, by exploiting the Gauss-Laguerre quadrature method \cite[Eq. (25.4.45)]{ref12}, a closed-form expression for \eqref{eq:2MGF} is obtained as

\begin{equation}
    C = \frac{1}{\ln(2)}\sum_{\ell=1}^{L} \mathcal{W}_\ell\frac{\mathcal{M}_Y(\mathcal{Z}_\ell)( 1 - \mathcal{M}_X(\mathcal{Z}_\ell))}{{\mathcal{Z}_\ell}},
\end{equation}
for borh \eqref{eq:Cq'} and \eqref{eq:Ce,q} the closed form is given by
\begin{equation}
C = \frac{1}{\ln(2)} \sum_{\ell=1}^{L} {\mathcal{W}_\ell} \frac{( 1 - \mathcal{M}_Y(Z_\ell))}{{\mathcal{Z}_\ell}},
\end{equation}
where \( \mathcal{W}_t \) and \( \mathcal{Z}_t \)  are respectively the weight factor of the Laguerre polynomial and the sample points, tabulated in \cite[Table. (25.9)]{ref12}.

\subsection*{B. Secrecy Rate }
The secrecy rates for each node $R_{q}^{{sec}} $ and $R_{q'}^{{sec}}$can be expressed as follows

%\begin{equation}
%R_{q}^{{sec}} = \left[C_{q} - \max C_{e,q}\right]^{+},
%\end{equation}
%\begin{equation}
%R_{q'}^{{sec}} = \left[C_{q'} - \max C_{e,q'}\right]^{+},
%\end{equation}
%where

  \vspace{-10pt}
\begin{equation}
\begin{split}
R_{q}^{{sec}} = \Bigg[ & \frac{1}{\ln(2)} \sum_{\ell=1}^{L} \frac{\mathcal{W}_\ell}{\mathcal{Z}_\ell} \left(1 + \frac{\mathcal{Z}_\ell \Omega_{URS} \rho_{s,q}}{m_{URS}}\right)^{-m_{URS}} \\
& \times \left(1 - \left(1 + \frac{\mathcal{Z}_\ell \Omega_{U'RS} \rho_{s,q'}}{m_{U'RS}}\right)^{-m_{U'RS}}\right) \\
& - \frac{1}{\ln(2)} \sum_{\ell=1}^{L} \mathcal{W}_\ell \frac{1 - \left(\frac{1}{1 + N \rho_{e,q} \mathcal{Z}_\ell}\right)}{{\mathcal{Z}_\ell}} \Bigg]^+,
\end{split}
\end{equation}
  \vspace{-10pt}
\begin{equation}
\begin{split}
R_{q'}^{{sec}} = \Bigg[ & \frac{1}{\ln(2)}\sum_{\ell=1}^{L} \mathcal{W}_\ell\frac{( 1 - (1+\frac{\mathcal{Z}_\ell\Omega_{U'RS}\rho_{s,q'}}{m_{U'RS}})^{-m_{U'RS}})}{{\mathcal{Z}_\ell}}\\
& - \frac{1}{\ln(2)}\sum_{\ell=1}^{L} \mathcal{W}_\ell\frac{1 -  (\frac{1}{1+N\rho_{e,q'} \mathcal{Z}_\ell})}{\mathcal{Z}_\ell} \Bigg]^{+}.
\end{split}
\end{equation}

Hence, the sum secrecy rate is obtained as:
\begin{equation}
    R_{{ sum}}^{{ sec}}=R_{q}^{{sec}}+R_{q'}^{{sec}}.
    \label{eq:counter}
\end{equation}

\section{Optimization of UAV Placement}

In this section, we optimize the UAV’s 2D positioning inside a given region and, given the positions of nodes, adopt a linear grid-based search algorithm where one coordinate is held fixed while searching for a sub-optimal value of the other coordinate in a sequential manner. This strategy ensures systematic convergence toward an optimal location that maximizes WSSR with reasonable computational complexity. The corresponding  problem is formulated in \textup{(\number\numexpr\getrefnumber{eq:counter}+1\relax)}, and the iterative algorithm to solve it is detailed in Algorithm~1.

\edef\baseeqnum{\number\numexpr\getrefnumber{eq:counter}+1\relax}

\begin{alignat*}{2}
  &\max_{C_u} \quad 
    w_1 R_{q}^{sec}+ w_2 R_{q'}^{sec}\tag{\baseeqnum a} \\
  &\quad\text{s.t.}
  \ \  x_{\min} \leq x_u \leq x_{\max}, && \tag{\baseeqnum b} \\
    &\quad \quad\quad  y_{\min} \leq y_u \leq y_{\max}, && \tag{\baseeqnum c} \\
&\quad \quad\quad0 \leq P_{q'},P_q \leq P_{\text{max}}, && \tag{\baseeqnum d} \\
&\quad  \ \ \ \ \ \   w_1 + w_2 = 1, w_i\in (0,1),  \, i \in \{1, 2\},&&
\tag{\baseeqnum e}\\
   &\quad\quad\quad\lambda_t^m + \lambda_r^m = 1,  \lambda_q^m \in (0, 1), \, q \in \{r, t\}, \, \forall m, \tag{\baseeqnum f}\\
\end{alignat*}

  \vspace{-10pt}

\begin{algorithm}[H]
\caption{Iterative Algorithm for Problem \textup{(\number\numexpr\getrefnumber{eq:counter}+1\relax)}}
\label{alg1_xy}
\begin{algorithmic}[1]
\STATE \textbf{Initialize:} UAV position $(x_u, y_u)$, where $x = x_{\min} : 1 : x_{\max}$, $y = y_{\min} : 1 : y_{\max}$. Set iteration index $k = 1$, and convergence threshold $\epsilon_0 > 0$.

\STATE Fix $y_u^{(k)}$, search over $x_u \in [x_{\min}, x_{\max}]$.

\STATE Evaluate the objective $R_{q}^{sec}+R_{q'}^{sec}$.

\STATE Select $x_u^{(k+1)}$ that maximizes the WSSR.

\STATE Fix $x_u^{(k+1)}$, search over $y_u \in [y_{\min}, y_{\max}]$.

\STATE Repeat the evaluation and update $y_u^{(k+1)}$.

\STATE Check convergence: $\|C_u^{(k+1)} - C_u^{(k)}\| < \epsilon_0$.

\STATE \textbf{break}

\STATE Set $k \leftarrow k + 1$ \textbf{Until} $\{k = K_{\max}\}$.

\STATE \textbf{Return}

\STATE $(x_u^*, y_u^*) =  C_u^{(k+1)}$
\end{algorithmic}
\end{algorithm}

\section{Numerical Results}
In this section, analytical results are validated through Monte-Carlo simulations and the numerical results are provided to evaluate the effectiveness of the proposed algorithms for the maximization of the WSSR of UAV-mounted STAR-RIS-assisted NOMA IoT system. We consider  AP to be located at  the origin of a Cartesian coordinate and a UAV-mounted STAR-RIS hovering over a specific region at fixed height of Z = 100m to reconfigure direct links between nodes and AP. The positions of nodes are assumed to be fixed in this region.

\begin{table}[htbp]
\centering
\caption{Simulation Parameters}
\label{table:sim_params}
\begin{tabular}{|>{\raggedright\arraybackslash}p{0.65\linewidth}|c|}
\hline
\textbf{Parameter Description} & \textbf{Value} \\\hline
Number of STAR-RIS elements $M$ & $40$\\\hline
 AP Coordinate $C_s$&$[0,0,10]$\\\hline
 UAV Coordinate $C_u$&$[50,50,100]$\\\hline
 Reflection Side Node Coordinate $U_r$&$[75 , -25,0 ]$\\\hline
 Transmission Side Node Coordinate $U_t$&$[125,75,0]$\\\hline
 Transmission Side malicious Node Coordinate $E_t$&$[25,75,0]$\\\hline
 Reflection Side malicious Node Coordinate $E_r$&$[50,25,0]$\\\hline
Maximum transmit power $P_{\max}$& $23$ dBm\\\hline
 Noise Power&$-100$ dBm\\\hline
The path loss exponent $a$& $2.7$\\\hline
 Path loss at 1 m $\beta_0$&$-20$ dB\\\hline
 Shape Parameters $m$&2\\\hline
 Spread Parameters $\Omega$&1\\\hline
Concentration parameter $\kappa$& 5\\\hline
Laguerre $L$& $30000$\\\hline
Convergence threshold $\epsilon_0, \epsilon_1, \epsilon_2$& $10^{-6}$\\ \hline
\end{tabular}
\end{table}

%\begin{figure}[htbp]
   % \centering
    %\includegraphics[width=0.5\textwidth]{positions.png} % Adjust the width as needed
    %\caption{positions of sccess point, UAV mounted STAR-RIS, nodes.}
    %\label{fig:positions of nodes of the system}
%\end{figure}

\begin{figure}[htbp]
   \vspace{-20pt} % Reduce space above
    \centering
    \includegraphics[width=0.4\textwidth]{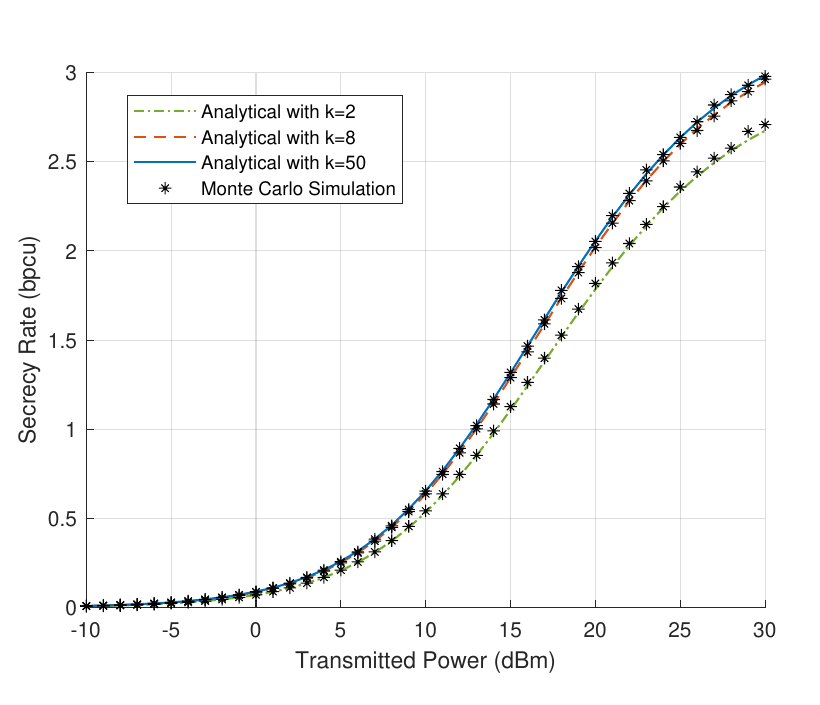} % Adjust the width as needed
    \vspace{-10pt} % Reduce space above
    \caption{Secrecy rate versus transmit power at the reflection side with $k \in \{2, 8, 50\}$.}
     \vspace{-20pt}
    
    \label{fig:reflection_secrecy_rate_vs_P}
\end{figure}

\begin{figure}[htbp]
   \vspace{-25pt} % Reduce space above
    \centering
    \includegraphics[width=0.4\textwidth]{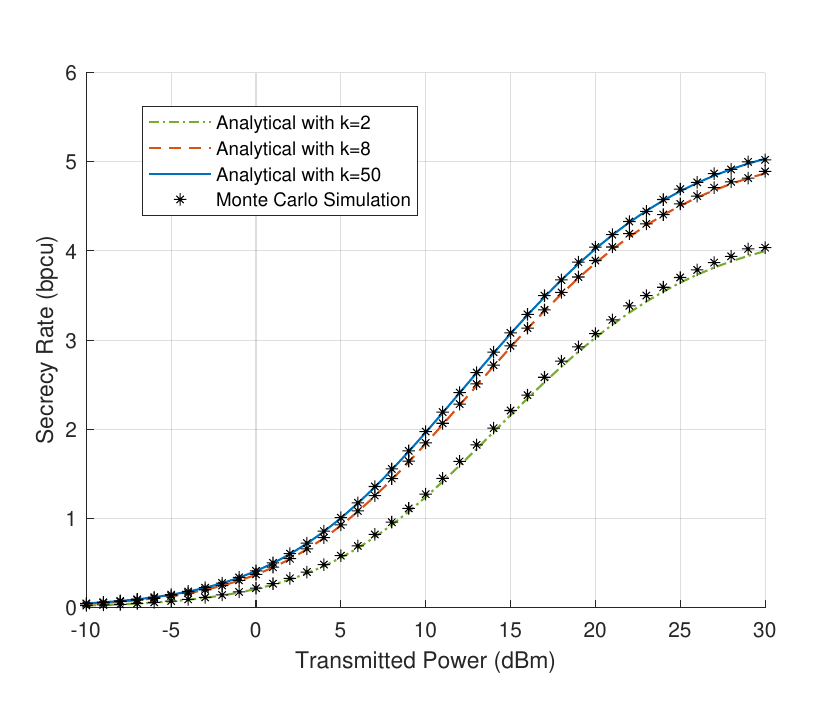} % Adjust the width as needed
    \vspace{-10pt} % Reduce space below
    \caption{Secrecy rate versus transmit power at the transmission side with $k \in \{2, 8, 50\}$.}
     \vspace{-10pt}
    \label{fig:transmission_secrecy_rate_vs_P}
\end{figure}

Fig.~\ref{fig:reflection_secrecy_rate_vs_P} and ~\ref{fig:transmission_secrecy_rate_vs_P} illustrate the secrecy rate performances at the reflection and transmission sides of the IoT network as a function in transmit power under different values of concentration parameter $k \in \{2, 8, 50\}$. The parameter $k$ describes the  phase estimation error of the reflection surface, as $k$ increases, the phase distribution becomes more concentrated, so less phase uncertainty. It can be observed that lower values of $k$ lead to a degradation in the  secrecy rate across all transmit power levels, and higher values of $k$ lead to very close curves. Moreover, the results show a close match between Monte Carlo simulations and the analytical expressions which validates the accuracy of the derivation. These observations highlight the critical impact of phase on the secrecy performance in STAR-RIS-assisted UAV IoT network.

%Fig.~\ref{fig:transmission_secrecy_rate_vs_P} shows the secrecy rate performance versus the transmit power for different values of the von Mises concentration parameter $k$, which characterizes the severity of phase errors in the system. As $k$ increases, the phase error decreases, leading to improved signal coherence and higher secrecy rates. It can be observed that the secrecy rate increases with both transmit power and $k$, verifying the sensitivity of secrecy performance to phase distortion. Moreover, the analytical results closely match the simulation results, which validates the accuracy of the derived expressions.

\begin{figure}[htbp]
  \vspace{-15pt} % Reduce space above 
    \centering
    \includegraphics[width=0.4\textwidth]{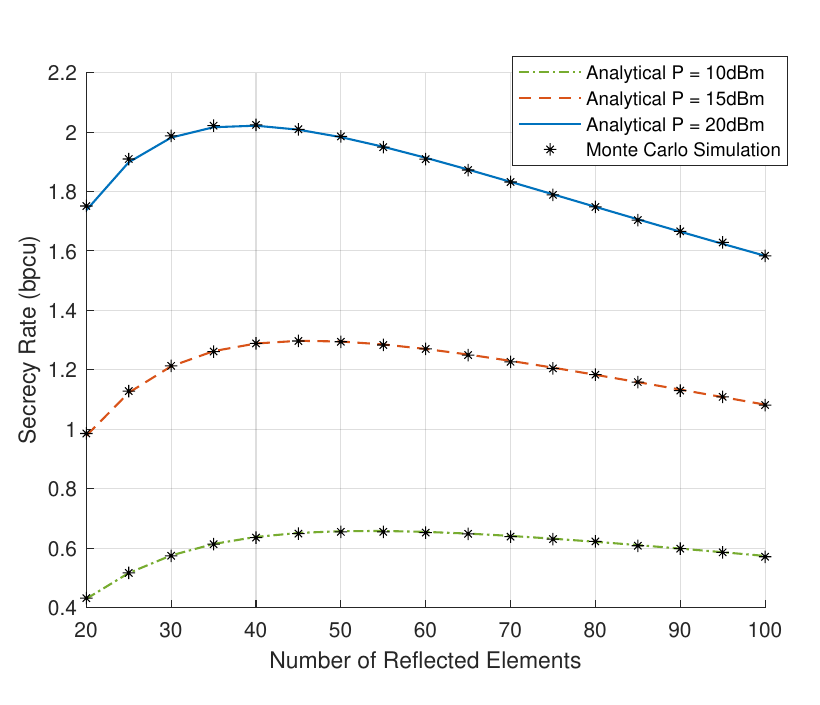} % Adjust the width as needed
    \vspace{-10pt} % Reduce space above
    \caption{Secrecy rate versus the number of reflected elements $M$ at the reflection side with $P \in \{10, 15, 20\mathrm{dBm}\}$}
    \label{fig:reflection_secrecy_rate_vs_M}
\end{figure}

\begin{figure}[htbp]
 \vspace{-10pt}
    \centering
    \includegraphics[width=0.4\textwidth]{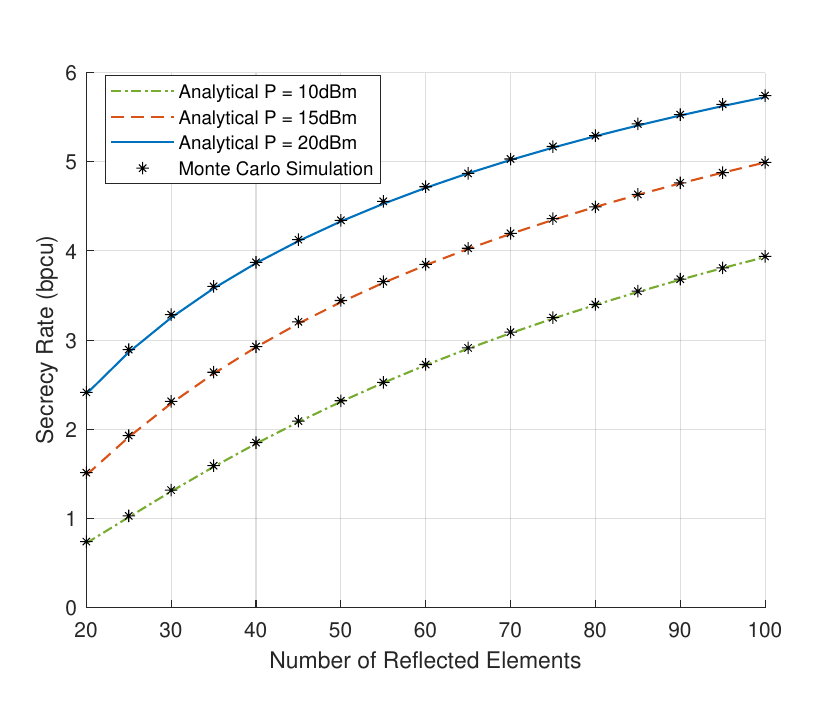} % Adjust the width as needed
     \vspace{-10pt}
    \caption{Secrecy rate versus the number of reflected elements $M$ at the transmission side with $P \in \{10, 15, 20\mathrm{dBm}\}$}
     \vspace{-10pt}
    \label{fig:transmission_secrecy_rate_vs_M}
\end{figure}

Fig.~\ref{fig:reflection_secrecy_rate_vs_M} shows that the secrecy performance of the reflection side of a UAV-mounted STAR-RIS improves with the number of reflected elements $M$ initially, due to enhanced beamforming gain. However, the secrecy rate starts to decrease beyond an optimal point (around $M = 40$), which is  a result of a quick increasing in eave rate with the number of reflecting elements. Increasing the transmit power level consistently yields higher secrecy rates, with $P = 20$\,dBm achieving the best performance.

Fig.~\ref{fig:transmission_secrecy_rate_vs_M} presents the secrecy rate performance at the transmission side of a UAV-mounted STAR-RIS system as a function of the number of reflected elements $M$, for different transmit power levels $P \in \{10, 15, 20\mathrm{dBm}\}$. Both Monte Carlo simulation results and analytical expressions are included. It is evident that the secrecy rate improves monotonically with increasing $M$, due to the enhanced beamforming gain provided by a larger number of STAR-RIS elements. Furthermore, higher transmit power levels leads to an enhanced secrecy rate, as expected. %The strong match between analytical and simulation results across all cases confirms the accuracy and robustness of the derived theoretical expressions.

%Fig.~\ref{fig:transmission_secrecy_rate_vs_M} illustrates the secrecy rate at the transmission side as a function of the number of reflecting elements $M$ for different transmit power levels. It is observed that the secrecy rate improves with increasing $M$, due to enhanced beamforming gain provided by a larger STAR-RIS aperture. In addition, higher transmit power contributes to an overall secrecy enhancement. The consistency between simulation and analytical results demonstrates the validity of the proposed theoretical analysis.

\begin{figure}[htbp]
 \vspace{-10pt}
    \centering
    \includegraphics[width=0.4\textwidth]{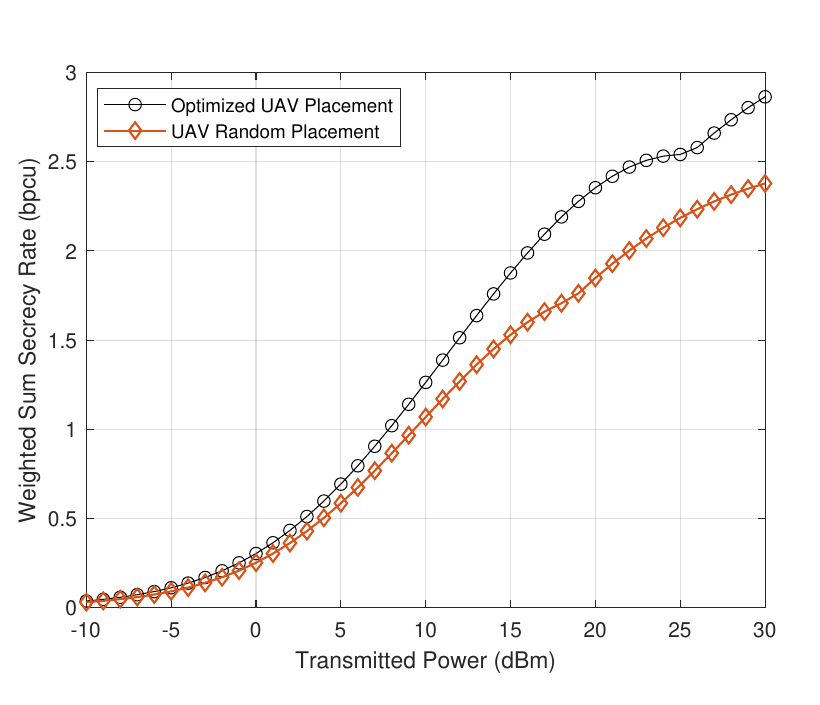} % Adjust the width as needed
     \vspace{-10pt}
    \caption{Impact of optimizing UAV placement and ES coefficient on  WSSR of the system.}
    \label{fig:optimized placement_vs_random placement}
\end{figure}

Fig.~\ref{fig:optimized placement_vs_random placement} illustrates the impact of the UAV placement optimization on the WSSR of the STAR-RIS-assisted uplink NOMA system when both nodes are transmitting with the same power, with $\lambda_t = 0.5, \lambda_r = 0.5$ and the weights of the secrecy rates are $w_1=0.5 , w_2=0.5$. It shows that optimizing the UAV placement results in a noticeable improvement in the WSSR across the entire range of transmit powers specially for high values. This improvement is attributed to the enhanced channel conditions and better interference mitigation facilitated by the optimized UAV location relative to nodes postions.

\begin{figure}[htbp]
    \centering
    \includegraphics[width=0.4\textwidth]{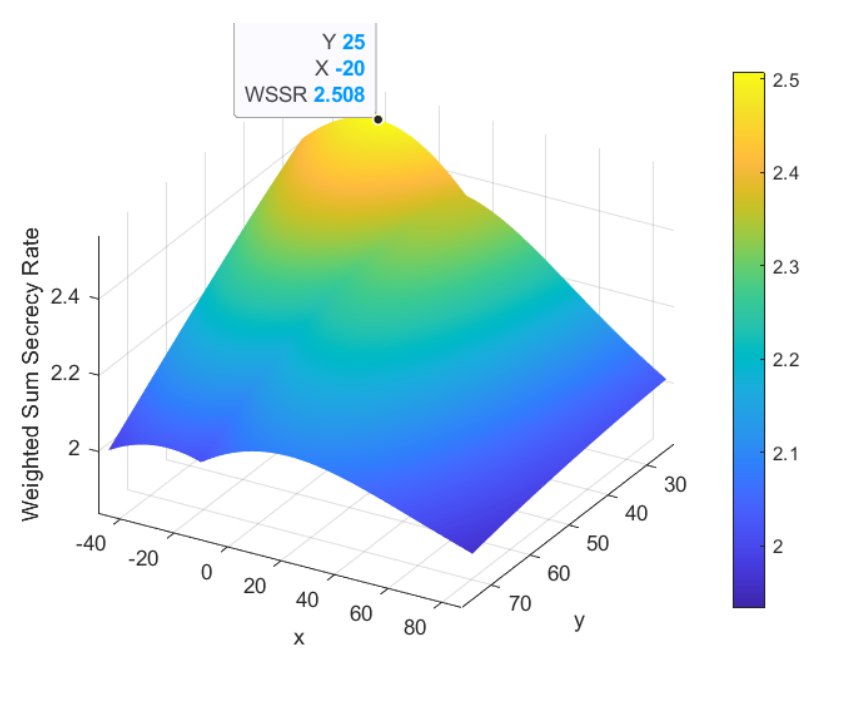} % Adjust the width as needed
     \vspace{-10pt}
    \caption{(WSSR) as a function of the UAV's horizontal position.}
    \label{fig:optimized UAV placement}
\end{figure}

 Fig.~\ref{fig:optimized UAV placement} shows the WSSR as a function of the UAV's horizontal coordinates $(x, y)$ in the proposed system. The surface reveals that the WSSR is highly sensitive to the UAV placement, with a well-defined maximum observed at approximately $x = -20$~m and $y = 25$~m, corresponding to a peak WSSR of $2.508$ bpcu. This highlights the critical role of optimal UAV deployment in enhancing secrecy performance. The results confirm that intelligently placing the UAV allows more effective exploitation of STAR-RIS capabilities, improving signal alignment and interference mitigation, thereby maximizing PLS in the uplink NOMA scenario.

\section{Conclusion}

This paper has investigated the secrecy performance of UAV-mounted STAR-RIS-assisted NOMA uplink systems under practical phase shift estimation errors caused by aerial platform imperfections. We derived closed-form expressions for the ergodic secrecy rates assuming the von Mises distribution to model phase errors, validated by Monte Carlo results. Our proposed UAV placement optimization effectively compensates for phase-induced performance degradation while maintaining security guarantees. The results demonstrate that optimal STAR-RIS configuration requires careful consideration of both spatial nodes distributions and UAV-induced phase error characteristics, particularly at higher transmit power levels. These results provide valuable design insights for practical implementations of secure aerial IoT systems employing STAR-RIS technology.

\section*{APPENDIX A}

Let the \(M\) complex random variables be defined as $|h_{U_q R_m}|\,|h_{R_m S}|\,e^{j\theta^{\mathrm{err}}_m}, \quad m = 1, \ldots, M$, where each variable is independent and identically distributed (i.i.d.). These variables have common first- and second-order statistics given by the mean \(\mu = \alpha^2 \varphi_1\), the variance \(\nu = 1 - \alpha^4|\varphi_1|^2\), and the pseudo-variance \(\rho = \varphi_2 - \alpha^4 \varphi_1^2\).

where \(\varphi_1, \varphi_2 \in \mathbb{R}\) are the first and second trigonometric moments of the phase error \(\theta^{\mathrm{err}}_m\),  and, consequently, \( \mu, \rho \in \mathbb{R} \). These expressions follow from the second-order characterization of complex random variables as detailed in~\cite{ref10}.

Under these assumptions, and by invoking the Central Limit Theorem, the sum of the \(M\) variables can be approximated as a complex normal distribution for large \(M\), i.e., \(\sqrt{A} \sim \mathcal{CN}\left(\mu, \frac{\nu}{M}, \frac{\rho}{M}\right)\).
Letting \(U = \Re\{\sqrt{A}\}\) and \(V = \Im\{\sqrt{A}\}\), the covariance between \(U\) and \(V\) is given by~\cite{ref10}:

\setcounter{equation}{\numexpr\getrefnumber{eq:counter} + 1\relax}

\begin{equation}
\mathrm{Cov}[U, V] = \frac{1}{2} \Im\left(-\frac{\nu}{M} + \frac{\rho}{M}\right) = 0. \label{eq:cov_uv}
\end{equation}

Since \(U\) and \(V\) are jointly Gaussian and uncorrelated, they are statistically independent. Furthermore, their distributions are given by \(U \sim \mathcal{N}(\mu, \sigma_U^2)\) and \(V \sim \mathcal{N}(0, \sigma_V^2)\), where:
\begin{equation}
\sigma_U^2 = \frac{1}{2} \Re\left(\frac{\nu}{M} + \frac{\rho}{M}\right) = \frac{1}{2M}(1 + \varphi_2 - 2\alpha^4\varphi_1^2), \label{eq:sigma_u}
\end{equation}
\begin{equation}
\sigma_V^2 = \frac{1}{2} \Re\left(\frac{\nu}{M} - \frac{\rho}{M}\right) = \frac{1}{2M}(1 - \varphi_2). \label{eq:sigma_v}
\end{equation}

\section*{APPENDIX B}

Consider the magnitude-squared variable $A = U^2 + V^2$.
Here, \(U^2/\sigma_U^2\) follows a non-central chi-squared distribution, while \(V^2\) follows a gamma distribution with shape parameter \(1/2\) and scale \(2\sigma_V^2\). Since \(U\) and \(V\) are independent (as shown in Appendix~A), the cumulant generating function (CGF) of \(A\), denoted \(K_A(t)\), is the sum of the individual CGFs:
\begin{equation}
\begin{aligned}
K_A(t) &= K_{U^2}(t) + K_{V^2}(t) \\
&= \frac{\mu^2 t}{1 - 2\sigma_U^2 t} - \frac{1}{2} \ln(1 - 2\sigma_U^2 t) - \frac{1}{2} \ln(1 - 2\sigma_V^2 t). \label{eq:KA}
\end{aligned}
\end{equation}

To simplify the expression for large \(M\), we expand the first term using a Maclaurin series:
\[
\frac{\mu^2 t}{1 - 2\sigma_U^2 t} = \mu^2 t + \mu^2 (2\sigma_U^2) t^2 + \mu^2 (2\sigma_U^2)^2 t^3 + \ldots
\]

\begin{equation}
       = \frac{\mu^2}{4\sigma_U^2} \sum_{k=1}^{\infty} \frac{k}{2^{k-1}} \frac{(4\sigma_U^2 t)^k}{k} = \frac{\mu^2}{4\sigma_U^2} \underbrace{\sum_{k=1}^{\infty} \frac{(4\sigma_U^2 t)^k}{k}}_{= -\ln(1-4\sigma_U^2 t)} - g(t) ,
\end{equation}
where  
\[
g(t) = \frac{\mu^2}{4\sigma_U^2} \sum_{k=3}^{\infty} \left( 1 - \frac{k}{2^{k-1}} \right) \frac{(4\sigma_U^2 t)^k}{k}
\]

\[
< \frac{\mu^2}{4\sigma_U^2} \sum_{k=3}^{\infty} \frac{(4\sigma_U^2 t)^k}{k} = 4\mu^2 \sigma_U^2 t^2 \sum_{k=1}^{\infty} \frac{(4\sigma_U^2 t)^k}{k+2}
\]

\[
< 4\mu^2 \sigma_U^2 t^2 \sum_{k=1}^{\infty} \frac{(4\sigma_U^2 t)^k}{k} = -4\mu^2 \sigma_U^2 t^2 \ln \left( 1 - 4\sigma_U^2 t \right).
\]

Given that \(\sigma_U^2 = \mathcal{O}(M^{-1})\), it follows that \(g(t) = \mathcal{O}(M^{-2})\). Therefore, the CGF \(K_A(t)\) can be approximated as
\begin{equation}
\begin{split}
K_A(t) &= -\frac{\mu^2}{4\sigma_U^2} \ln(1 - 4\sigma_U^2 t) - \frac{1}{2} \ln(1 - 2\sigma_U^2 t) \\
&\quad - \frac{1}{2} \ln(1 - 2\sigma_V^2 t). \label{eq:KA2}
\end{split}
\end{equation}

This expression represents the CGF of the sum of three independent gamma-distributed random variables with parameters \((\frac{\mu^2}{4\sigma_U^2}, 4\sigma_U^2)\), \((\frac{1}{2}, 2\sigma_U^2)\), and \((\frac{1}{2}, 2\sigma_V^2)\), respectively.  While the resulting distribution can be characterized based on \cite{ref11}, we approximate \(K_A(t)\) by:
\begin{equation}
K_A(t) = -\frac{\mu^2}{4\sigma_U^2} \ln(1 - 4\sigma_U^2 t), \label{eq:KA3}
\end{equation}
which introduces an error of order \(\mathcal{O}(M^{-1})\). Under this approximation, \(A\) follows a gamma distribution with shape parameter \(\frac{\mu^2}{4\sigma_U^2}\) and scale parameter \(4\sigma_U^2\).

As a result, \(\sqrt{A}\) follows a Nakagami distribution with fading (shape) parameter
\begin{equation}
m = \frac{\mu^2}{4\sigma_U^2} = \frac{M}{2} \cdot \frac{\alpha^4 \varphi_1^2}{1 + \varphi_2 - 2\alpha^4 \varphi_1^2}, \label{eq:m_nakagami}
\end{equation}
and spread parameter \(\mu^2\).

\section*{Acknowledgment}

This work was supported and funded by the Deanship of Scientific Research at Imam Mohammad Ibn Saud Islamic University (IMSIU) (grant number: IMSIU-DDRSP2504).


\begin{thebibliography}{00}

\bibitem{ref1} 
B. Li, Z. Fei, and Y. Zhang, ``UAV communications for 5G and beyond: Recent advances and future trends,'' \textit{IEEE Internet Things J.}, vol. 6, no. 2, pp. 2241--2263, Apr. 2019.

\bibitem{ref2} 
B. Zheng, C. You, W. Mei, and R. Zhang, ``A survey on channel estimation and practical passive beamforming design for intelligent reflecting surface aided wireless communications,'' \textit{IEEE Commun. Surv. Tutor.}, vol. 24, no. 2, pp. 1035--1071, 2nd Quart., 2022.

\bibitem{ref3} 
Z. Zhai, X. Dai, B. Duo, X. Wang, and X. Yuan, ``Energy-efficient UAV-mounted RIS assisted mobile edge computing,'' \textit{IEEE Wireless Commun. Lett.}, vol. 11, no. 12, pp. 2507--2511, Dec. 2022.

\bibitem{ref4} 
X. Liu, Y. Yu, B. Peng, X. B. Zhai, Q. Zhu, and V. C. M. Leung, ``RIS-UAV enabled worst-case downlink secrecy rate maximization for mobile vehicles,'' \textit{IEEE Trans. Veh. Technol.}, vol. 72, no. 5, pp. 6129--6141, May 2023.

\bibitem{ref5} 
X. Mu, Y. Liu, L. Guo, J. Lin, and R. Schober, ``Simultaneously transmitting and reflecting (STAR) RIS aided wireless communications,'' \textit{IEEE Trans. Wireless Commun.}, vol. 21, no. 5, pp. 3083--3098, May 2022.

\bibitem{ref6} 
Y. Liu \textit{et al.}, ``STAR: Simultaneous transmission and reflection for 360° coverage by intelligent surfaces,'' \textit{IEEE Wireless Commun.}, vol. 28, no. 6, pp. 102--109, Dec. 2021.

\bibitem{16}
K. Rabie \textit{et al.}, ``On the performance of UAV-mounted reconfigurable intelligent surfaces,'' \textit{IEEE Commun. Lett.}, vol. 25, no. 10, pp. 3285--3289, Oct. 2021.

\bibitem{20}
Q. Wu and R. Zhang, ``Joint active and passive beamforming optimization for intelligent reflecting surface assisted UAV communications,'' \textit{IEEE Trans. Wireless Commun.}, vol. 18, no. 11, pp. 5394--5409, Nov. 2019.

\bibitem{18}
R. Long \textit{et al.}, ``Reflections and beyond: Intelligent reflecting surface aided wireless networks,'' \textit{IEEE Trans. Wireless Commun.}, vol. 20, no. 1, pp. 420--431, Jan. 2021.

\bibitem{19}
X. Mu \textit{et al.}, ``Intelligent reflecting surface enhanced UAV communication,'' \textit{IEEE Trans. Veh. Technol.}, vol. 70, no. 10, pp. 10104--10109, Oct. 2021.

\bibitem{36}
H. Chen \textit{et al.}, ``Phase error modeling and performance analysis for STAR-RIS-aided systems,'' \textit{IEEE Trans. Commun.}, vol. 70, no. 12, pp. 8063--8077, Dec. 2022.

\bibitem{39}
X. Zhang \textit{et al.}, ``Statistical beamforming design for STAR-RIS aided secure communication under channel uncertainties,'' \textit{IEEE Trans. Inf. Forensics Security}, vol. 17, pp. 765--778, 2022.

\bibitem{zhang2022star}
Z.~Zhang, J.~Chen, Y.~Liu, Q.~Wu, B.~He, and L.~Yang, ``On the secrecy design of STAR-RIS assisted uplink NOMA networks,'' \emph{IEEE Trans. Wireless Commun.}, vol.~21, no.~12, pp.~11207--11221, Dec. 2022.

\bibitem{li2023secrecy}
X.~Li, Y.~Zheng, M.~Zeng, Y.~Liu, and O.~A.~Dobre, ``Enhancing secrecy performance for STAR-RIS NOMA networks,'' \emph{IEEE Trans. Veh. Technol.}, vol.~72, no.~2, pp.~2684--2688, Feb. 2023.

%\bibitem{han2022an}
%Y.~Han, N.~Li, Y.~Liu, T.~Zhang, and X.~Tao, ``Artificial noise aided secure NOMA communications in STAR-RIS networks,'' \emph{IEEE Wireless Commun. Lett.}, vol.~11, no.~6, pp.~1191--1195, Jun. 2022.

\bibitem{tang2021ris}
Z.~Tang, T.~Hou, Y.~Liu, J.~Zhang, and C.~Zhong, ``A novel design of RIS for enhancing the physical layer security for RIS-aided NOMA networks,'' \emph{IEEE Wireless Commun. Lett.}, vol.~10, no.~11, pp.~2398--2401, Nov. 2021.

\bibitem{wang2025covert}
Q.~Wang, S.~Guo, C.~Wu, C.~Xing, N.~Zhao, D.~Niyato, and G.~K.~Karagiannidis, ``STAR-RIS aided covert communication in UAV air-ground networks,'' \emph{IEEE J. Sel. Areas Commun.}, vol.~43, no.~2, pp.~1--1, Jan. 2025.

\bibitem{salem2024phase}
A.~Salem, K.-K.~Wong, and C.-B.~Chae, ``Impact of phase-shift error on the secrecy performance of uplink RIS communication systems,'' \emph{IEEE Trans. Wireless Commun.}, vol.~23, no.~7, Jul. 2024.

\bibitem{li2024active}
X.~Li, Y.~Pei, X.~Yue, Y.~Liu, and Z.~Ding, ``Secure communication of active RIS assisted NOMA networks,'' \emph{IEEE Trans. Wireless Commun.}, early access, 2024.

\bibitem{8}
J. D. Vega Sanchez, P. Ramirez-Espinosa, and F. J. Lopez-Martinez, ``Physical layer security of large reflecting surface aided communications with phase errors,'' \textit{IEEE Wireless Commun. Lett.}, vol. 10, no. 2, pp. 325--329, Feb. 2021.

\bibitem{9}
D. Diao \textit{et al.}, ``Reflecting elements analysis for secure and energy-efficient UAV-RIS system with phase errors,'' \textit{IEEE Wireless Commun. Lett.}, vol. 13, no. 2, pp. 293--297, Feb. 2024.

\bibitem{10}
J. Sun \textit{et al.}, ``Leveraging UAV-RIS reflects to improve the security performance of wireless network systems,'' \textit{IEEE Netw. Lett.}, vol. 5, no. 3, pp. 104--107, Sep. 2023.

\bibitem{ref9} 
K. A. Hamdi, ``A useful lemma for capacity analysis of fading interference channels,'' \textit{IEEE Trans. Commun.}, vol. 58, no. 2, pp. 411--416, Feb. 2010.

\bibitem{ref12}
M. Abramowitz and I. A. Stegun, \emph{Handbook of Mathematical Functions with Formulas, Graphs, and Mathematical Tables}, 10th ed. New York, NY, USA: Dover, 1972.

\bibitem{ref10} 
B. Picinbono, ``Second-order complex random vectors and normal distributions,'' \textit{IEEE Trans. Signal Process.}, vol. 44, no. 10, pp. 2637--2640, Oct. 1996.

\bibitem{ref11} 
P. G. Moschopoulos, ``The distribution of the sum of independent gamma random variables,'' \textit{Ann. Inst. Stat. Math.}, vol. 37, no. 3, pp. 541--544, Dec. 1985.

\end{thebibliography}
\end{document}